# Theoretical study of the dual harmonic system and its application on the CSNS/RCS

Yao-Shuo Yuan, Na Wang, Shou-Yan Xu, Yue Yuan, and Sheng Wang

Dongguan branch, Institute of High Energy Physics, CAS, Guangdong Province, Dongguan 523803, China

Corresponding Author: Sheng Wang

Email address: wangs@ihep.ac.cn

Postal address: No.1, Zhongziyuan Road, 523803, Dalang Town, Dongguan, Guangdong Province, China

Telephone: +86-10-88235976

Supported by National Natural Science Foundation of China (11175193)

**Abstract**

The dual harmonic system has been widely used in high intensity proton synchrotrons to suppress the space charge effect, as well as reduce the beam loss. To investigate the longitudinal beam dynamics in the dual rf system, the potential well, the sub-buckets in the bunch and the multi-solutions of the phase equation have been studied theoretically. Based on these theoretical studis, the optimization of bunching factor and rf voltage waveform are made for the dual harmonic rf system in the upgrade phase of the CSNS/RCS. In the optimization process, the simulation with space charge effect is done by using a newly developed code C-SCSIM.

**1 Introduction**

Space charge effect is the principal cause of emittance growth and beam loss in high intensity accelerators. In high intensity synchrotron, the space charge effect is the limit of the maximum number of particles that can be accumulated. The space charge induced tune shift is always used as a standard of space charge, and in a synchrotron, the relation between the bunching factor $B_f$ and the tune shift $\Delta \nu$ is [1]

$$\Delta \nu = - \frac{r_p n_t}{2\pi \beta^2 \gamma^3 \varepsilon B_f}, \qquad (1)$$

where $r_p$ is the classic radius of proton, $n_t$ is the bunch population, $\varepsilon$ is the transverse emittance, $\beta$ and $\gamma$ are the relativistic factors. To decrease the space charge tune shift, increasing the bunching factor by using dual harmonic rf system is a common method in the high intensity proton synchrotron [2-5].

China Spallation Neutron Source (CSNS) [6-7] is a pulsed neutron source with beam power of 100 kW, which is now under construction. Its accelerator consists of an 80 MeV H- linac and a 1.6 GeV proton Rapid Cycling Synchrotron (RCS). In the future upgrade, the beam power will upgrade to 500 kW. As an important way of depressing the space charge effect, the dual harmonic rf system will be employed in RCS in the upgrade stage. In this paper, some important beam dynamics issues in the dual rf system, such as the potential well, the multi-solutions of the phase equation and the sub-bucket in the bunch has been theoretically studied. Based on these theoretical study, the optimization of bunching factor and rf voltage waveform are made for the dual harmonic rf system in the upgrade stage of the CSNS/RCS. In the optimization process, the simulation with space charge effect is done by using a newly developed code C-SCSIM [8]. The relation between bunching factor and the rf voltage has been studied and the corresponding beam distribution in the phase space is presented.

## 2. The theoretical study

### 2.1 The potential well

To obtain the formula for the potential well concisely, coordinate (ф, P) is used [9]. Here, $\phi$ is the phase relative to the fundamental rf cavity and

$$P = -h|\eta|\delta/v_s \quad (2)$$

where $h$ is the ratio between the harmonic numbers in fundamental rf system and higher harmonic rf system (h=2 for dual rf system), $\eta$ the phase slip factor, $\delta$ the momentum deviation, $v_s$ the synchrotron tune at zero amplitude for the fundamental rf system. The Hamiltonian in this coordinate system can be written as

$$H(P) = \frac{1}{2}v_s P^2 + U_1(\phi) + U_2(\phi) \quad (3)$$

$$U_1(\phi) = v_s[\cos\phi_{1s} - \cos\phi + (\phi_{1s} - \phi)\sin\phi_{1s}] \quad (4)$$

$$U_2(\phi) = \frac{v_s r}{h}[\cos\phi_{2s} - \cos[\phi_{2s} + h(\phi - \phi_{1s})] - h(\phi - \phi_{1s})\sin\phi_{2s}] \quad (5)$$

where $U_1$ and $\phi_{1s}$ are the potential energy and synchronous phase in the fundamental cavity, while $U_2$ and $\phi_{2s}$ are for the second-harmonic cavity, $r$ is the ratio of the second-harmonic voltage and

fundamental voltage, i.e. $r = -V_2/V_1$. The total voltage can be expressed as

$$V(\phi) = V_1 \sin(\phi) + V_2 \sin(2\phi - \phi_2) \quad (6)$$

where $\phi_2$ is the rf phase angle of the second-harmonic cavity relative to the fundamental one,

$$\phi_2 = 2\phi_{1s} - \phi_{2s} \quad (7)$$

Substituting Eq.(7) into Eq.(5), the potential energy in the dual harmonic rf system can be expressed as

$$U(\phi) = v_s \left\{ (-\cos\phi - \phi\sin\phi_s) + r[\frac{1}{2}\cos(2\phi - \phi_2) + \phi\sin(2\phi_s - \phi_2)] \right\} \quad (8)$$

**2.2 The effect of $r$ on the potential wall**

For simplicity, we assume $v_s$ equals 1. The curves of the function $U(\phi)$ in Eq.(8) are plotted in Fig.(1), which shows that the shape of the potential well changes as $r$ varies from 0.4 to 0.8. From the figure we can see that when $r$ becomes larger, the bottom of the potential well becomes higher.

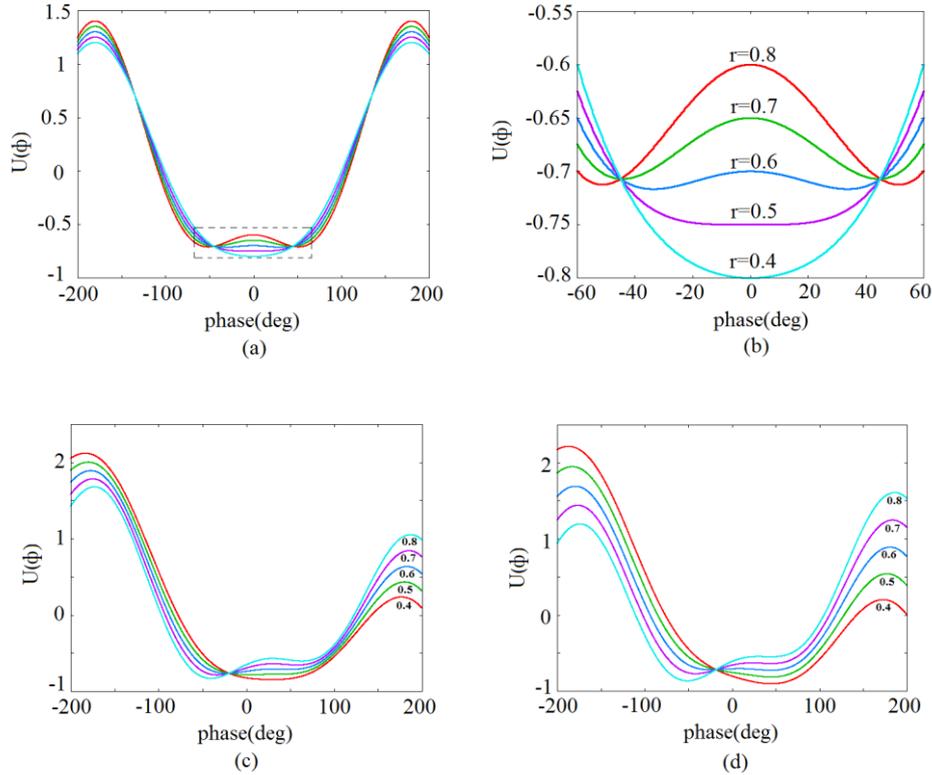

Figure 1: The variation of the potential energy with different $r$. (a) $\phi_s=\phi_2=0$; (b) enlarged view of (a) in the bottom; (c) $\phi_s=30$ deg; $\phi_2=30$deg; (d) $\phi_s=45$deg; $\phi_2=15$deg

In order to explain the results, we take the first-order derivative of the potential formula in

Eq.(8):

$$\frac{dU(\phi)}{d\phi} = v_s \{(\sin\phi - \sin\phi_s) - r[\sin(2\phi - \phi_s) - \sin(2\phi_s - \phi_2)]\} \quad (9)$$

which gives,

$$\left.\frac{dU(\phi)}{d\phi}\right|_{\phi=\phi_s} \equiv 0 \quad (10)$$

Therefore, the slope at $\phi=\phi_s$ identically equals to zero, and is independent of the value of $\phi_s$ and $\phi_2$. Moreover, the second derivative is

$$\frac{d^2U(\phi)}{d\phi^2} = \cos\phi - 2r\cos(2\phi - \phi_s) \quad (11)$$

Let $r$ equal to 0.5, we have

$$\left.\frac{d^2U(\phi)}{d\phi^2}\right|_{\phi=\phi_s} \equiv 0 \quad (12)$$

which means the potential well is always flat at $\phi=\phi_s$.

**2.3 Multi-solutions and two sub-buckets**

If we assume $v_s=1$, $\phi_s=0$ and $\phi_2=0$, Eq.(8) can be simplified to

$$\frac{dU(\phi)}{d\phi} = \frac{\omega_0 eV_1}{2\pi\beta^2 E}[\sin\phi - r\sin(2\phi)] = 0, \quad (13)$$

which gives

$$2r\cos\phi - 1 = 0 \quad (14)$$

The relation between $r$ and $\phi$ in Eq.(14) are listed in Table 1. It can be seen that when r>0.5, Eq.(8) has two solutions, corresponding to the two bottoms of potential well, as shown in Fig.(1)(a)(b).

Table 1: The relation between $r$ and $\phi$

| r | ϕ(deg) |
|---|---|
| 0.4 | Nan |
| 0.5 | 0 |
| 0.6 | ±33.6 |
| 0.7 | ±44.4 |
| 0.8 | ±51.3 |

On the other hand, the total rf voltage in the dual harmonic system is

$$V(\phi) = V_1[\sin(\phi) - rV_2 \sin(2\phi - \phi_2)] \tag{15}$$

In the accelerating process, the rf voltage should be

$$V(\phi) = \rho L \frac{dB(t)}{dt} \tag{16}$$

where $\rho$ and $L$ denote the bending radius and the circumstance of the ring, $B$ is the magnetic field of the dipole. As shown in Fig.(2), when r<0.5, Eq.(16) has only one solution, i.e. the synchronous phase $\phi_s$ but when r>0.5, there are three solutions: one is the synchronous phase $\phi_s$, and other two solutions are $\phi_{s1}$ and $\phi_{s2}$. When $r>0.5$, the voltage function is no longer monotonic.

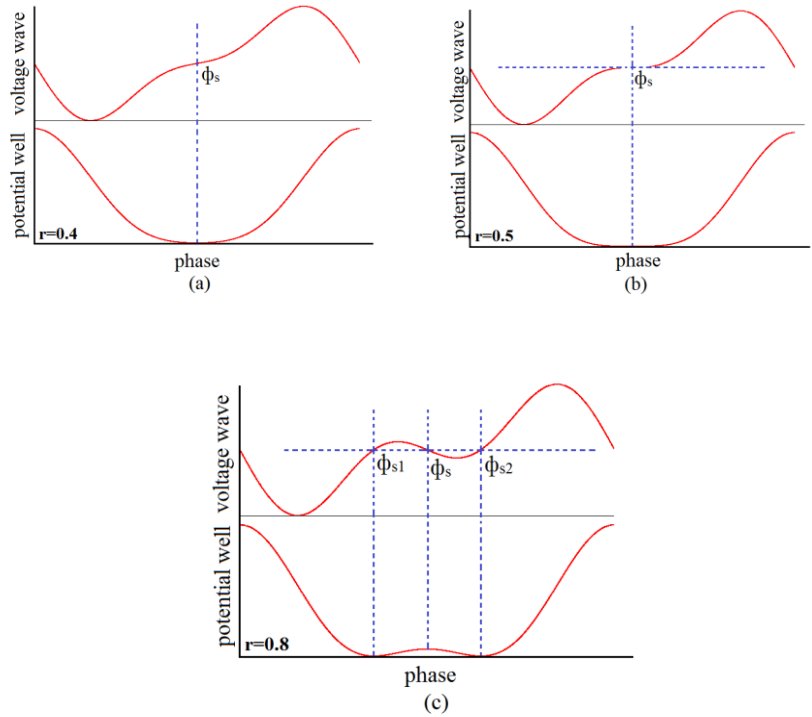

Figure 2: Schematic drawing of the voltage wave and the potential at (a) r=0.4; (b) r=0.5 and (c) r=0.8

In which, $\phi_{s1}$ and $\phi_{s2}$ are corresponding to the two bottoms of the potential well. In this case, particles in the bucket can be divided into two parts by their energy deviation ΔE, those with small ΔE are trapped within the two "depression" in the potential well, and can only oscillate within the range of the two sub-buckets, as marked by rectangular dot in Fig.(3); those with larger ΔE can extend to all the whole bucket, as marked by round dot. The "sub-bucket" plotted as a schematic drawing in Fig.(3), can be observed by beam simulation, as shown in Fig.(5-6).

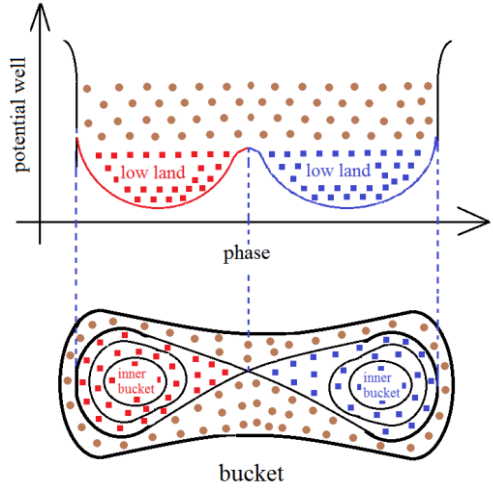

Figure 3: The schematic drawing of particles with various ΔE in the potential well and in the two sub-buckets

### 2.4. The effects of $\phi_s$ and $\phi_2$ on the potential well

Let r=0.5, the curve of the potential well varies with different $\phi_s$ and $\phi_2$ are shown in Fig.(4).

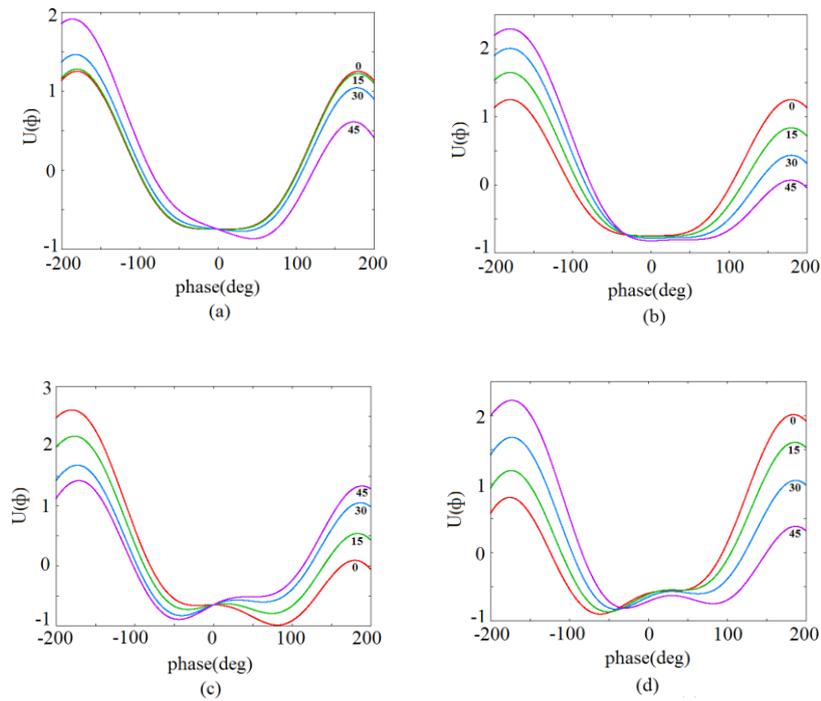

Figure 4: the curve of the potential well with r=0.5: (a) $\phi_s$ vs. U, with $\phi_2$=0; (b) $\phi_2$ vs. U, with $\phi_s$=0; (c) $\phi_s$ vs U, with $\phi_2$=30°; (d) $\phi_2$ vs. U, with $\phi_s$=30°

To sum up, the *r* affects the extent of the flatness in the bottom of the potential well, which is the reason of the formation of the sub-bucket in a bucket. The bottom is symmetrical about the line ϕ=ϕ$_s$ when ϕ$_s$=ϕ$_2$=0, and becomes uneven when either ϕ$_s$≠0 or ϕ$_2$≠0. The uneven depends on the value of $\phi_s$ or $\phi_2$.

## 3 The application on the CSNS/RCS

### 3.1 The optimization of bunching factor

In section 2, we have obtained theoretically the formula of potential well and how the four coefficients $\phi_s$, $\phi_2$, $V_1$ and $V_2$ influence the shape of the potential well. In fact, theses parameters are not independent. Also it is difficult to find out an analytic solution for the bunching factor, which often can be considered as a crucial parameter in the dual harmonic system and taken as a key criterion in the rf voltage optimization process.

In the upgrade stage of CSNS/RCS, the dual harmonic system will be employed to achieve uniform longitudinal beam distribution (i.e. the flat potential well) and much larger bunching factor. Early study has been performed for the upgrade case of a 200 kW beam power [10]. Here, by using the simulation code C-SCSIM, the relation between the potential well and the bunching factor for the case of 500 kW beam power has been investigated. Table 2 lists the input parameters used in the simulation.

Table 2: Main input parameters used in the simulation

| Circumference(m) | 227.92 |
| :---: | :---: |
| Bending radius(m) | 8.021 |
| Repetition rate(Hz) | 25 |
| Harmonic number | 2 |
| Chopping factor | 50% |
| Injection turn | 500 |
| Injection energy(MeV) | 250 |
| Extraction energy(GeV) | 1.6 |
| Number of protons($10^{13}$) | 7.8 |
| Starting time at injection(ms) | -0.3 |
| Macro-particle for injection | 20000 |

The simulation has been performed in two steps. First, during 0-2ms in an accelerating process, the value of $r$ ($r=-V_2/V_1$) is scanned from 0.4 to 0.9 by increasing $V_2$ while keeping $V_1$ as a constant in each time point, as listed in Table 3. The simulation results are shown in Fig.(5). One can see that as *r* increases, both the top and the bottom of the bucket shrink while the bunching factor increases at first and then decreases. Moreover, when r>0.5, the bunch divides into two sub-buckets, and they trend to become more and more "clear" as *r* increases. We can also observe that when *r* equals 0.5, the bucket becomes flat, but the bunching factor is not at its maximum.

Table 3: The fundamental voltage waveform used in step 1 of the simulation

| Time(ms) | Fundamental voltage(kV) |
|---|---|
| 0 | 50 |
| 1 | 95 |
| 2 | 110 |

Secondly, let the $V_1$ and $V_2$ change, but keep *r* as a constant. Let the value of $V_1$ and $V_2$ used in the Fig.5(f) be a reference, change the value to its 0.6, 0.8, 1.2 and 1.4 times. The simulation results are shown in Fig.(6), from which we can observe that as $V_1$ and $V_2$ increase, the area of the bucket is enlarged but the area of the bunch and the bunching factor does not change much compared with the case in Fig.(5).

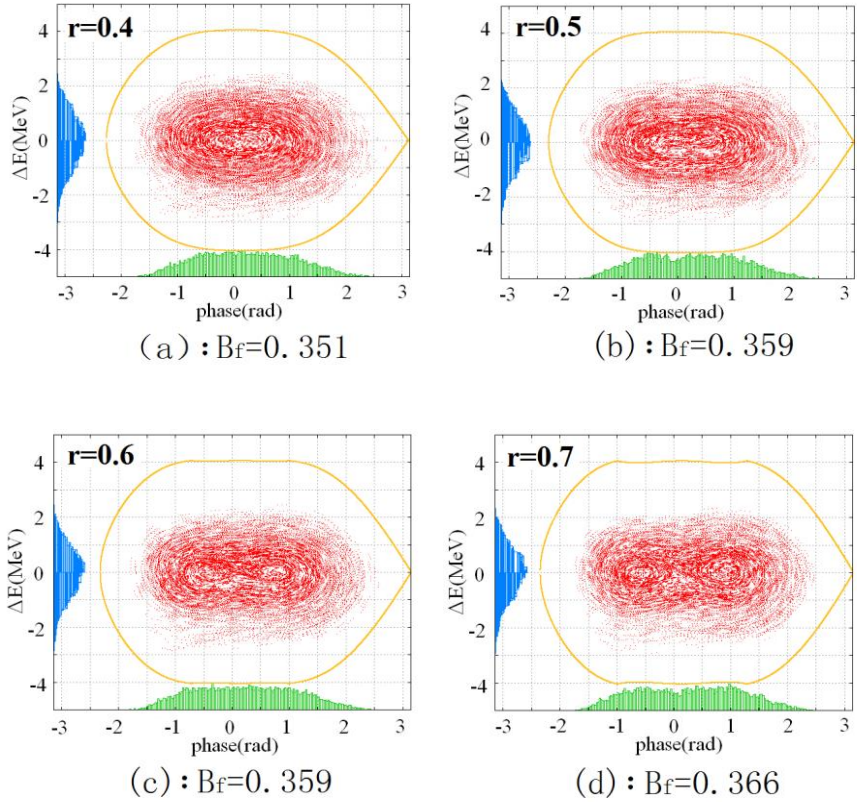

(a): Bf=0.351
(b): Bf=0.359
(c): Bf=0.359
(d): Bf=0.366

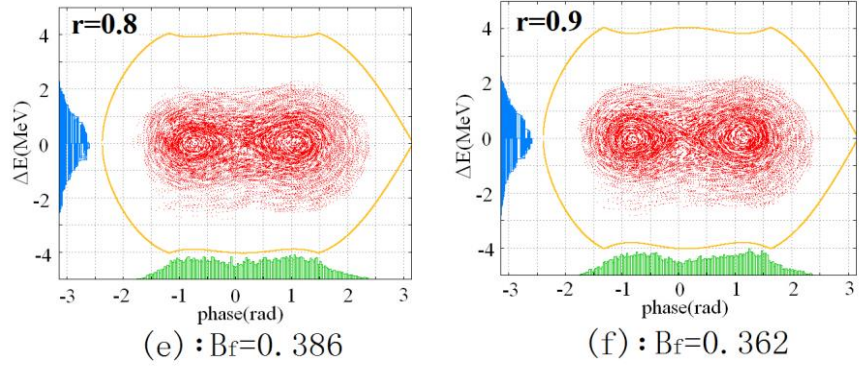

Figure 5: the particle distribution with different *r* at the end of the injection process (500[th] turn)

（ϕ$_s$=8.2deg，ϕ$_2$=14.0deg, V$_1$=77.8kV）

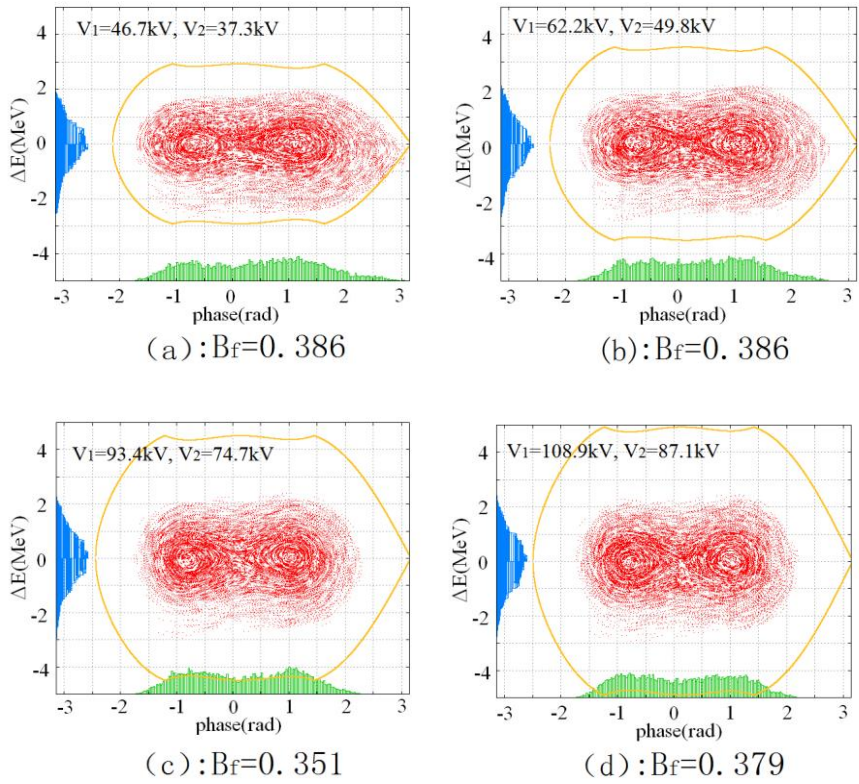

Figure 6: The particle distributions with different *V$_1$* and *V$_2$* at the end of the injection process

### 3.2 The optimization of rf voltage waveform

According to the property of the bunching factor obtained above, the voltage waveform can be optimized with an iteration method using the code C-SCSIM. Generally, the optimization procedure consists of two steps. At first, the fundamental rf voltage are calculated using the iteration method [3]. Secondly, the second rf voltage waveform are optimized according to the beam distribution and the bunching factor calculated by using the code.

The calculated fundamental rf voltage waveform and the optimized second-harmonic rf voltage waveform are shown in Fig.7. Fig.8 gives the particle distribution at 500th turn for the upgrading CSNS/RCS, using only the calculated fundamental rf voltage waveform and the dual harmonic system, respectively. From Fig.8(b) it can be seen that the longitudinal beam emittance and the bunching factor is 1.5 eVs and 0.51 respectively, which meets the design requirement.

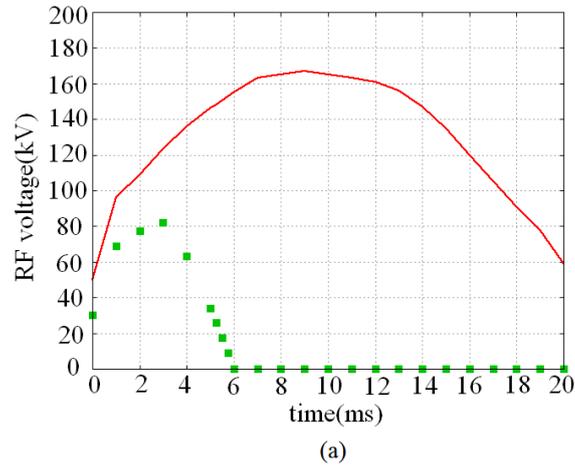

(a)

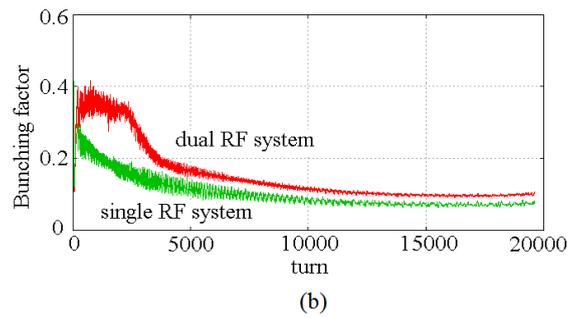

(b)

Figure 7: (a) the optimized voltage waveform; (b) the comparison of the bunching factor in dual and single rf system

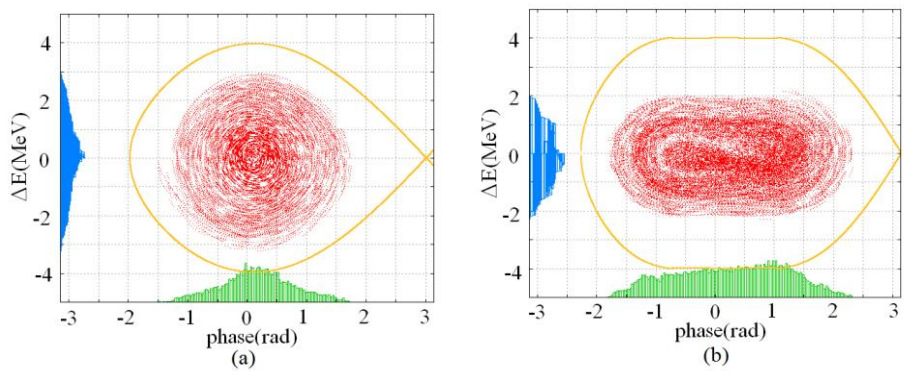

Figure 8: the comparison of the beam distribution at 500th turn for upgrading phase with single rf system and dual rf system

**4 Summary**

Beam dynamics for the dual harmonic rf system has been investigated by theoretical analysis. The influence of the rf voltage, the relative phase $\phi_2$ and the synchronous phase $\phi_s$ on the shape of the potential well has been introduced. The bunching factor and the formation and the characteristic of the sub-buckets in the bunch have been illustrated. These theoretical results are applied on the optimization design of dual harmonic rf system for the future upgrade of CSNS/RCS, and the results is important for the dual rf system design in the upgrade stage.